\documentstyle[multicol,epsf,aps]{revtex}
\begin{document}
\draft

\title{The inclusive $^{56}$Fe($\nu_e,e^-$)$^{56}$Co cross section}

\author{E.Kolbe$^{1,}$\footnote{Permanent address: Departement f\"ur
    Physik und Astronomie der Universit\"at Basel, Basel,
    Switzerland}, K. Langanke$^2$ and G. Mart\'{\i}nez-Pinedo$^2$}

\address{$^1$Institut f\"ur Kernphysik I, Forschungszentrum Karlsruhe,
  Germany \\
  $^2$Institut for Fysik og Astronomi, {\AA}rhus Universitet, DK-8000
  {\AA}rhus C, Denmark} 

\date{\today} 
\maketitle

\begin{abstract}
  We study the $^{56}$Fe($\nu_e,e^-)$$^{56}$Co cross section for the
  KARMEN neutrino spectrum. The Gamow-Teller contribution to the cross
  section is calculated within the shell model, while the forbidden
  transitions are evaluated within the continuum random phase
  approximation. We find a total cross section of $2.73 \times
  10^{-40}$ cm$^2$, in agreement with the data.
\end{abstract}


\begin{multicols}{2}
  
The search for neutrino oscillations is one of the most promising
experimental pathways for possible physics beyond the standard model
of electroweak interaction. KARMEN is one of the pioneering low-energy
accelerator-based experiments searching for these modes.  It consists
of liquid scintillator housed in a 7000 ton shielding blockhouse
\cite{Zeitnitz}.  One of the experimental signals which the KARMEN
experiment can observe is the energy distribution of the electron
emitted in charged-current reactions.  This gives the experiment also
the capability to obtain nuclear structure informations, and in fact
the KARMEN collaboration has measured several (inclusive and
exclusive) neutrino-induced charged and neutral current cross sections
on $^{12}$C \cite{Maschuw}, which is an important ingredient of the
liquid scintillator. Recently the collaboration has also been able to
determine inclusive $(\nu_e, e^-)$ cross sections for $^{13}$C and
$^{56}$Fe. The latter might be quite important for two reasons: First,
it provides a benchmark to test the ability of nuclear models to
describe neutrino-induced reactions on nuclei in the iron mass range
as they are encountered in astrophysical scenarios like the supernova
collapse \cite{Woosley}.  Second, future long baseline neutrino
oscillation experiments like MINOS \cite{Minos} or OMNIS \cite{Omnis}
will use steel as detector material. Thus a rather reliable knowledge
of the charged-current (and neutral-current) neutrino-induced cross
section on $^{56}$Fe is desirable to support the potential ability to
observe supernova neutrinos by this detector.

In this communication we present the first calculation (apart from a
shell model estimate by Bugaev et al.\cite{Bu79}) of the
$^{56}$Fe($\nu_e,e^-)$$^{56}$Co cross section.  The charged-current
neutrino-induced cross section for a transition from a nuclear state
with angular momentum $J$ and isospin quantum numbers $T, M_T$ to a
final nuclear state ($J',T'M'_T$) is derived in Refs.
\cite{Walecka,Donnelly}. Its evaluation reduces to the calculation of
the various vector and axial vector multipole operators between
appropriate nuclear many-body states. To derive these states we have
adopted the following strategy. In Ref. \cite{Kolbe99} it has been
demonstrated that the inclusive cross section for forbidden
transitions is rather well described within the continuum random phase
approximation (RPA).  This model is, however, less reliable for the
allowed transitions which in the long wavelength approximation are
mediated by the Gamow-Teller operator $\bbox{\sigma \tau}$.  Here, the
continuum RPA does not recover all nucleon-nucleon correlations
necessary to reproduce the quenching of the Gamow-Teller strength.
Such a task can be achieved in large-scale shell model calculations
performed in the complete pf shell. On the other hand, at the higher
neutrino energies present in the KARMEN $\nu_e$-spectrum a replacement
of the $\lambda=1^+$ multipole operator by $\bbox{\sigma \tau}$ is not
quite appropriate. However, considering the momentum-transfer
dependence of the operator within shell model approaches represents a
very formidable computational effort, even on modern parallel
computers. To reliably estimate the contribution of the $\lambda=1^+$
operator to the cross section, we have therefore chosen a hybrid
approach of shell model and RPA, as discussed below.

Our shell-model calculation of the Gamow-Teller part to the
$^{56}$Fe($\nu_e,e^-)$$^{56}$Co cross section uses the
Strassburg-Madrid codes ANTOINE \cite{Caurier89} and NATHAN
\cite{Caurier99a} developed by Caurier. The calculation has been
performed within the complete pf shell adopting the recently developed
modified KB3 interaction. In this force slight monopole deficiencies
encountered in the original KB3 interaction have been corrected.  Ref.
\cite{Caurier99} demonstrates that the modified KB3 version gives a
fair account of nuclear spectra in the mass range $A=41-65$ and
reproduces the Gamow-Teller strength distributions in general very
well, if the latter are scaled by the factor $(0.74)^2$ which accounts
for the fact that complete $0 \hbar \omega$ shell-model calculations
overestimate the experimental GT strength by a universal factor
\cite{Wildenthal88,Langanke95,Martinez96}. We will adopt in the
following the same universal scaling factor.  With the code NATHAN it
is now possible to calculate the total GT strength for $^{56}$Fe in
the complete pf shell involving 7~413488 $J^\pi=0^+$ configurations
(this corresponds to an m-scheme dimension of 501 million).  We obtain
a total GT$_-$ strength (this is the direction in which a neutron is
changed into a proton) of 8.85 units which agrees with the
experimental value of $9.9\pm2.4$ \cite{Rapaport}. However, a
calculation of the GT$_-$ strength distribution $S_{GT}(E)$, where $E$
denotes the excitation energy in the final nucleus $^{56}$Co, in the
complete pf shell is still yet beyond present-day computer abilities.
We have therefore truncated the model space in our calculation of the
GT$_-$ strength distribution of $^{56}$Fe allowing a maximum of 5
particles to be excited from the $f_{7/2}$ orbital to the rest of the
pf shell in the final nucleus. Furthermore the Ikeda sum rule has been
fulfilled within the model space. The m-scheme dimension of this
calculation is 1.6 million for $^{56}$Fe and 19.8 million for
$^{56}$Co.  In our truncated calculation we obtain a total GT$_-$
strength of 9.3 units, showing that the calculation at this level of
truncation is nearly converged.

Due to the energy dependence of the phase space, the cross section is
sensitive to the Gamow-Teller strength distribution rather than only
to the total strength. Our shell-model calculation reproduces the
experimental GT$_-$ strength distribution quite well, as can be seen
in Fig.~2 of Ref.~\cite{Caurier99}.

For the Gamow-Teller transition we then have the total cross section
\begin{equation}
   \sigma(E_\nu) = \frac{G_F^2 \cos^2 \theta_c}{\pi} \int p_e E_e
   F(Z+1,E_e) S_{GT}(E) dE_e \; \; ,
\end{equation}
where $E_\nu$ is the neutrino energy, $G_F$ is the Fermi constant,
$E_e$ and $p_e$ are the energy and momentum of the outgoing electron,
respectively, $\theta_c$ the Cabibbo angle, $E = E_\nu - E_e$ due to
energy conservation, and $F(Z+1,E_e)$ the Fermi function which
accounts for Coulomb distortion of the outgoing electron wave
function.

In the KARMEN experiment the neutrino beam has a Michel energy
spectrum,
\begin{equation}
n(E_\nu) = \frac{96 E_\nu^2}{M_\mu^4} (M_\mu - 2 E_\nu) \; \; ,
\end{equation}
where $M_\mu$ is the muon mass. Folding Eq. (1) with the spectrum (2)
yields a partial $^{56}$Fe($\nu_e,e^-)$$^{56}$Co cross section of
139$\times 10^{-42}$ cm$^2$ for the Gamow-Teller transitions.  As
mentioned above, our treatment of the $\lambda=1^+$ multipole is only
correct for small neutrino energies. To estimate the effect introduced
by replacement of the $\lambda=1^+$ operator by the momentum-transfer
independent Gamow-Teller we have performed two calculations within the
random phase approximation. In the first we adopted the complete
$\lambda=1^+$ multipole operator \cite{Walecka}, in the second we
replaced this operator by $\bbox{\sigma \tau}$.  We then find partial
$^{56}$Fe($\nu_e,e^-)$$^{56}$Co cross sections of 341$\times 10^{-42}$
cm$^2$ and 431$\times 10^{-42}$ cm$^2$, respectively.  From these
calculations we conclude that the replacement of the full
$\lambda=1^+$ operator by the Gamow-Teller operator increases the
cross section by about $20\%$. If we correct our shell model estimate
by this factor, we derive at our final result for the partial
$^{56}$Fe($\nu_e,e^-)$$^{56}$Co cross section of 110$\times 10^{-42}$
cm$^2$.

The contributions of the other transitions to the cross section have
been calculated using the continuum RPA.  For nuclei with closed-shell
configurations this model is well documented in the
literature~\cite{Kolbe92}. We have recently generalized this formalism
to allow also for partial occupancies in the parent ground state
\cite{Kolbe99}. From our shell-model calculation we find the following
occupation numbers for the pf shell orbitals: $n_{7/2}=5.30$,
$n_{3/2}=0.32$, $n_{1/2}=0.07$ and $n_{5/2}=0.30$ for protons
$n_{7/2}=7.29$, $n_{3/2}=1.57$, $n_{1/2}=0.44$ and $n_{5/2}=0.70$ for
neutrons.  We assumed a completely occupied $^{40}$Ca core. The
single-particle energies were derived from an appropriate Woods-Saxon
potential. We used the experimental values for the proton and neutron
separation energies.  As the residual interaction we adopted the
finite-range G-matrix derived from the Bonn potential \cite{Speth}.

The other allowed transition is mediated by the $0^+$ operator, which
reduces to the Fermi transition in the limit of vanishing momentum
transfer.  The Fermi matrix element is easily obtained from isospin
symmetry, yielding a total Fermi strength of $S_F=(N-Z)$ which is
concentrated in the isobaric analog state at 3.5 MeV in $^{56}$Co. The
Fermi contribution to the $^{56}$Fe($\nu_e,e^-)$$^{56}$Co cross
section is then readily calculated within the RPA approach. We find a
partial cross section of 53$\times 10^{-42}$ cm$^2$.

The partial $^{56}$Fe($\nu_e,e^-)$$^{56}$Co cross sections for
selected forbidden transitions are listed in Table~\ref{tab:tab1}.
From the forbidden transitions only the $J^\pi=1^-$ and $2^-$ dipole
multipoles are important. We find a summed cross section of $111
\times 10^{-42}$ cm$^2$ for the forbidden transitions, which is only
slightly smaller than the contributions stemming from the allowed
transitions.

\begin{figure}
  \begin{center}
    \leavevmode
    \epsfxsize=0.8\columnwidth
    \epsffile{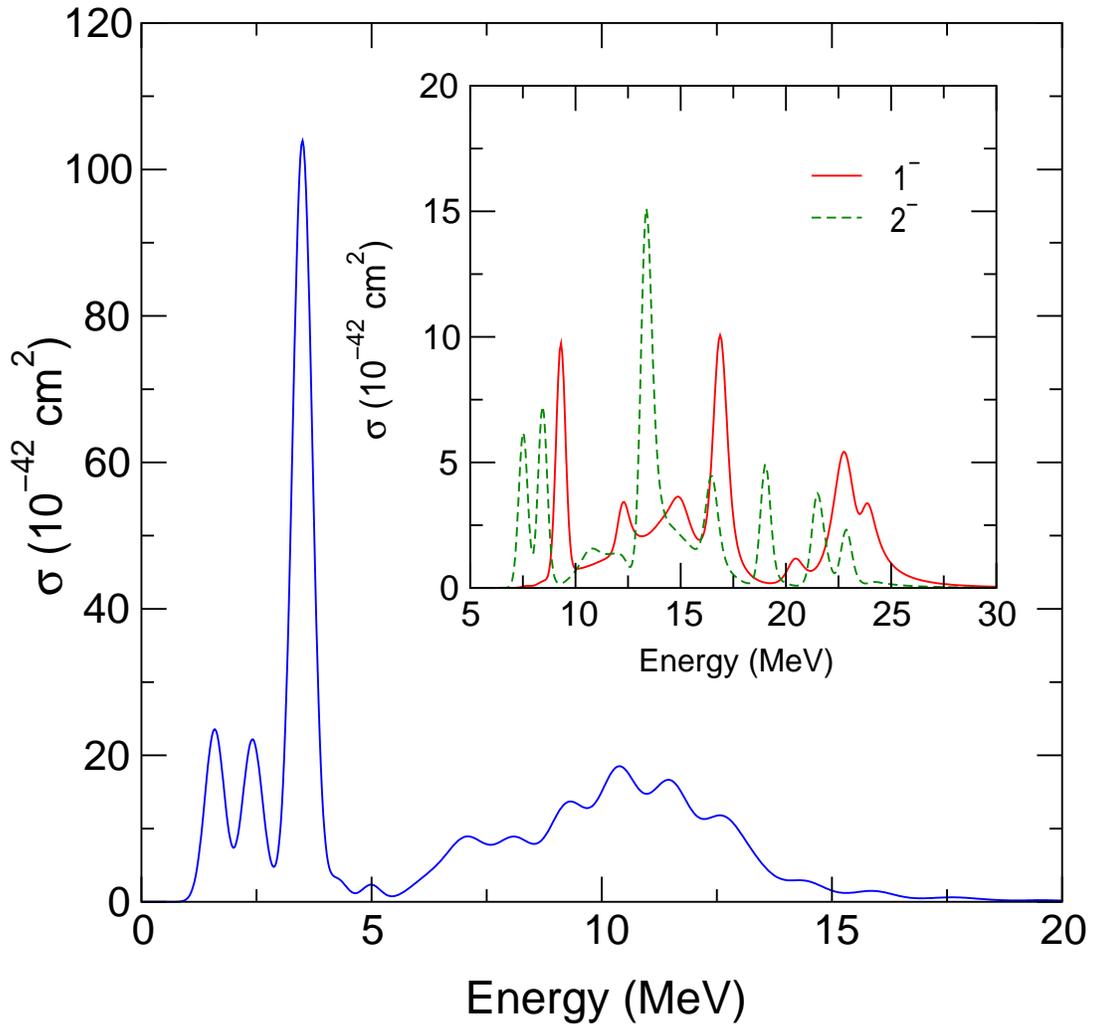}
    \caption{Differential $^{56}$Fe($\nu_e,e^-)$$^{56}$Co cross section
      for the KARMEN neutrino spectrum as function of excitation
      energy in $^{56}$Co.  The figure shows the allowed
      contributions, while the insert gives the contributions of the
      $1^-$ and $2^-$ multipolarities. The allowed contributions have
      been folded with a Gaussian of 0.5 MeV FHWM at energies below 5
      MeV and with 1 MeV FHWM above 5 MeV.}
    \label{fig:fig1}
    \end{center}
\end{figure}

Adding up the allowed and forbidden transitions, we find a total
$^{56}$Fe($\nu_e,e^-)$$^{56}$Co cross section of 2.73 $\times
10^{-40}$ cm$^2$, which agrees with the KARMEN result,
$2.56\pm1.08({\rm stat.})\pm0.43({\rm syst.}) \times 10^{-40}$ cm$^2$
\cite{Eitel}. Fig.~\ref{fig:fig1} shows the differential
$^{56}$Fe($\nu_e,e^-)$$^{56}$Co cross section as function of
excitation energy in $^{56}$Co.

To test the importance of the partial occupancy formalism for the
forbidden transitions, we have repeated the continuum RPA calculation
with a parent ground state whose occupation numbers have been taken
from the independent-particle model.  Thus, $n_{7/2}=6$ for protons,
and $n_{7/2}=8$ and $n_{3/2}=2$ for neutrons. The respective partial
cross sections are also listed in Table~\ref{tab:tab1}.  Confirming
the results and arguments given in Ref.~\cite{Kolbe99}, we observe
that an improved description of the parent ground state, i.e. by
considering partial occupancies, does not change the results
noticeably. We now find a total cross section of $ 2.62 \times
10^{-40}$ cm$^2$.

In summary, we have calculated the $^{56}$Fe($\nu_e,e^-)$$^{56}$Co
cross section for a neutrino spectrum corresponding to the KARMEN
experiment. While the Fermi contribution for the transition to the
isobaric analog state is trivial, we calculate the Gamow-Teller piece
to the cross section within a shell model approach which has been
proven to correctly describe the GT$_-$ strength distribution and
total strength for $^{56}$Fe. For the forbidden transitions we
employed the continuum RPA.  Combining all relevant multipoles we find
a total $^{56}$Fe($\nu_e,e^-)$$^{56}$Co cross section for the KARMEN
neutrino spectrum of $2.73 \times 10^{-40}$ cm$^2$ which agrees with
the experimental value obtained by the KARMEN collaboration,
$256\pm108({\rm stat.})\pm43({\rm syst.}) \times 10^{-42}$. Despite
the rather large experimental uncertainty, this agreement is promising
and shows that with the currently chosen combination of nuclear models
(shell model for allowed transitions and continuum RPA for forbidden
transitions) one should be able to reliably evaluate neutrino-induced
cross sections on nuclei in the iron mass range as they are of
interest in supernova simulations. Such calculations are in progress.

\acknowledgements

We thank K. Eitel for supplying us with the newest KARMEN cross
sections. We are grateful to Petr Vogel for useful comments on the
manuscript. Our work was supported in part by the Danish Research
Council.

\end{multicols}

\begin{table}
  \begin{center}
    \caption{Partial cross sections $\sigma_J^{\pi}$ for the
      $^{56}$Fe($\nu_e,e^-$)$^{56}$Co reaction induced by
      muon-decay-at-rest neutrinos.  The second and third columns give
      the (continuum RPA) results calculated without and with the
      consideration of partial occupancies in the ground state. The
      Gamow-Teller result ($J^\pi=1^+$) has been calculated within the
      shell model. The $J=0^+$ multipole contribution reflects the
      Fermi transition to the isobaric analog state. The cross
      sections are in units of $10^{-42}$ cm$^2$.}
    \label{tab:tab1}
    \begin{tabular}{ccc}
      multipole & $\sigma_J^{\pi}$ & $\sigma_J^{\pi}$ \\
      \hline
      $1^+$     &    110.6         &  109.8  \\
      $0^+$     &     45.7         &   52.7  \\
      \hline
      $0^-$     &      0.7         &    0.7  \\
      $1^-$     &     48.2         &   48.9  \\
      $2^+$     &      6.7         &    6.9  \\
      $2^-$     &     43.6         &   47.2  \\
      $3^+$     &      6.8         &    6.4  \\
      $3^-$     &      0.4         &    0.5  \\
      \hline
      $\Sigma$   &    262.7         &   273.1
    \end{tabular}
  \end{center}
\end{table}

\end{document}